\documentclass[%
reprint,
superscriptaddress,
 amsmath,amssymb,
 aps,
]{revtex4-2}

\usepackage{graphicx}
\usepackage{dcolumn}
\usepackage{bm}
\usepackage[hidelinks]{hyperref}
\usepackage{color}
\usepackage{siunitx}
\usepackage{physics}

\usepackage{pifont}
\usepackage{placeins}

\begin{document}

\preprint{APS/123-QED}

\title{
Theoretical Proposal of a Digital Closed-Loop Thermal Atomic-Beam Interferometer for High-Bandwidth, Wide-Dynamic-Range, and Simultaneous \\
Absolute Acceleration–Rotation Sensing
}

\author{Tomoya Sato}
\affiliation{Institute of Integrated Research, Institute of Science Tokyo,\\
4259 Nagatsuta-cho, Midori-ku, Yokohama, Kanagawa, 226-8501, Japan}

\author{Toshiyuki Hosoya}
\affiliation{Product Development Center, Japan Aviation Electronics Industry, Ltd.,\\
3-1-1, Musashino, Akishima-shi, Tokyo, 196-8555, Japan}

\author{Martin Miranda}
\affiliation{Institute of Integrated Research, Institute of Science Tokyo,\\
4259 Nagatsuta-cho, Midori-ku, Yokohama, Kanagawa, 226-8501, Japan}

\author{Hiroki Matsui}
\affiliation{Institute of Integrated Research, Institute of Science Tokyo,\\
4259 Nagatsuta-cho, Midori-ku, Yokohama, Kanagawa, 226-8501, Japan}

\author{Yuki Miyazawa}
\affiliation{Institute of Integrated Research, Institute of Science Tokyo,\\
4259 Nagatsuta-cho, Midori-ku, Yokohama, Kanagawa, 226-8501, Japan}

\author{Mikio Kozuma}
\email{kozuma@qnav.iir.isct.ac.jp}
\affiliation{Institute of Integrated Research, Institute of Science Tokyo,\\
4259 Nagatsuta-cho, Midori-ku, Yokohama, Kanagawa, 226-8501, Japan}
\affiliation{Department of Physics, Institute of Science Tokyo,\\
2-12-10 Ookayama, Meguro-ku, Tokyo, 152-8550, Japan}

\date{\today}

\begin{abstract}
We present a theoretical proposal and simulation study of a digital closed-loop thermal atomic-beam interferometer for inertial navigation applications. The scheme synchronizes phase biasing with momentum-kick reversal through the atomic transit time, extracting four interferometric phases to suppress Raman beam path-length errors, while two-photon detuning feedback maintains a pseudo-inertial frame and eliminates cross-coupling. The interferometer enables simultaneous measurements of acceleration and rotation based on an absolute, atom-interferometric reference, with high bandwidth and a wide dynamic range.
Numerical simulations verify that acceleration and angular velocity can be measured simultaneously and independently in real time without cross-coupling, demonstrating the absolute, decoupled nature of the proposed measurement scheme.
We further evaluate the noise-limited performance of the sensor and obtain sensitivities of \SI{3}{\micro m / s^2 / \sqrt{Hz}} (velocity random walk) and \SI{15}{\micro deg / \sqrt{h}} (angular random walk) for a \SI{170}{\degreeCelsius} $^{85}$Rb beam and an interferometer arm length of 100~mm, surpassing the performance of sensors currently used in state-of-the-art inertial navigation systems.
\end{abstract}

\maketitle

\section{Introduction}
Atom interferometers, similar to their optical counterparts, can detect angular velocities via the Sagnac effect, while finite mass in atoms make them also sensitive to accelerations.
The high sensitivity resulting from the extremely short de Broglie wavelength of matter waves enables exceptional performance in the field of precision measurement~\cite{hoganLightpulseAtomInterferometry2009, fangAdvancesAtomicGyroscopes2012, narducciAdvancesFieldableAtom2022}.
In angular velocity measurements, they have now surpassed the performance of state-of-the-art conventional gyroscopes~\cite{savoieInterleavedAtomInterferometry2018a}.
In absolute gravity measurements, they are already replacing classical macroscopic sensors~\cite{antoni-micollierDetectingVolcanoRelatedUnderground2022}, while in gravity-gradient measurements they have demonstrated subsurface structure monitoring~\cite{strayQuantumSensingGravity2022a}.
Looking ahead, they hold great promise for next-generation fundamental physics experiments, including tests of general relativity~\cite{biedermannTestingGravityColdatom2015}, gravitational-wave detection~\cite{dimopoulosAtomicGravitationalWave2008}, and searches for dark-sector physics~\cite{abeMatterwaveAtomicGradiometer2021}.

Inertial navigation, which estimates position without relying on external references~\cite{tittertonStrapdownInertialNavigation2004}, is a key application area for atom-interferometer-based inertial sensors—particularly in GPS-denied environments such as underwater or underground.
Such applications require the simultaneous measurement of both acceleration and angular velocity, combining high precision with wide bandwidth and large dynamic range.
Classical inertial navigation systems achieve this by hybridizing multiple types of sensors~\cite{el-sheimyInertialSensorsTechnologies2020}, for example, pairing optical gyroscopes with mechanical accelerometers.
A key advantage of matter-wave sensors is their ability to obtain both quantities with a single device, reducing sensor alignment errors and enabling fully self-contained inertial navigation.
Furthermore, the well-defined optical transition frequencies of the atoms provide an absolute physical reference, ensuring intrinsic scale-factor calibration and reducing long-term drift compared with conventional sensors.

State-of-the-art cold-atom experiments have already demonstrated six-axis inertial sensing using three mutually orthogonal pairs of counter-propagating Raman beams~\cite{canuelSixAxisInertialSensor2006}.
By exploiting low atomic velocities and a large-diameter Raman beam, time-multiplexed pulse sequences from a single Raman laser form the interferometers for all axes, inherently suppressing sensitivity to arbitrary laser phase, thereby permitting the simultaneous and independent determination of acceleration and angular velocity.
While this cold-atom approach has proven highly successful, systems based on laser-cooled atoms are typically limited to response bandwidths of only a few hertz to several tens of hertz.
Inertial navigation, however, generally requires several hundred hertz, motivating the use of thermal atomic beams.
In addition, the higher-order responses of atom interferometers to acceleration and angular velocity become more pronounced at lower atomic velocities~\cite{hoganLightpulseAtomInterferometry2009,bongsHighorderInertialPhase2006} and can be suppressed by using thermal atomic-beam implementations, further enhancing their suitability for such applications.

There are several challenges that must be addressed to meet the requirements of inertial navigation with thermal atomic beam interferometers.
Due to the short interaction time arising from the high longitudinal velocity, these systems require three spatially separated Raman beam pairs, rather than a temporally separated single Raman beam pair as in cold-atom setups.
In this configuration, the measurement becomes sensitive to arbitrary laser phase differences between Raman beams.
Although the elimination of laser-phase contributions and the measurement of the absolute value of angular velocity can still be achieved using counter-propagating atomic-beam interferometers~\cite{gustavsonRotationSensingDual2000}, absolute acceleration cannot be determined.
Furthermore, thermal-atom interferometers generally suffer from a limited dynamic range due to the broad longitudinal velocity distribution of atoms.
While our previous study~\cite{satoClosedloopMeasurementsAtominterferometer2025} successfully mitigated this limitation for angular velocity measurements by controlling the two-photon detuning of the Raman beams in a closed-loop manner,
only partial recovery of the measurement contrast has been achieved for acceleration, and the absolute value remains inaccessible.

In this paper, we propose a new operation scheme for a Mach–Zehnder interferometer in the spatial domain, referred to as the “digital closed-loop” method, to meet the performance requirements of inertial navigation.
This approach is inspired by the digital closed-loop concept that has been indispensable for achieving high precision and wide dynamic range in fiber-optic gyroscopes (FOGs).
In the digital closed-loop of FOGs~\cite{lefevreFiberOpticGyroscope2022}, digital signal processing of the sensor's analog output is taken as a given, and the laser phase is biased in an alternating manner using the intrinsic propagation time of light through the fiber coil as the unit cycle.
The interferometer output phase is determined directly from the photodetector current at positive and negative phase shifts, as opposed to being extracted via lock-in demodulation with a reference signal.
By performing multi-step phase-biasing operations, a variety of experimental parameters as well as physical information related to the Sagnac effect can be extracted and used to construct the closed loop.
This approach, for example, enables the detection of temporal variations in the output of the phase modulator and feedback to the appropriate value~\cite{lefevreFiberOpticGyroscope2022,lefevreOpticalfiberMeasuringDevice1992}, thereby naturally suppressing the long-term drift of the FOG. Furthermore, by applying serrodyne modulation with a step size corresponding to the angular velocity to be canceled, the system can achieve response at its intrinsic maximum speed.
This feature is critically important for inertial navigation, because if the responses of the three sensing axes are not tightly synchronized, the vehicle’s three-dimensional rotation cannot be estimated correctly, leading to large navigation errors.
Through this closed-loop operation, the FOG remains in a pseudo-inertial frame, thereby maintaining maximum sensitivity and eliminating various nonlinear responses.
The system can continue operating even when the Sagnac phase induced by angular velocity exceeds $2\pi$, thus achieving a higher dynamic range compared with the open-loop configuration.

In our proposed digital closed-loop method for atom interferometers, one of the Raman lasers is alternately biased with positive and negative phase shifts at intervals corresponding to the mean transit time of atoms through the interferometer, analogous to the digital closed-loop operation used in FOGs, enabling direct readout of the interferometric signal from the atomic fluorescence intensity.
The momentum kick from the Raman transition is reversed ($k$-reversal~\cite{durfeeLongTermStabilityAreaReversible2006, kwolekNoiseResilienceHighBandwidth2025}) after each complete set of positive and negative bias shifts, yielding four interferometer output phases.
These phases are used to determine the optical path length differences among the three Raman beams, which are then actively driven to zero via feedback, eliminating their influence on acceleration and angular velocity measurements.

During each $k$-reversal cycle, closed-loop operation is realized by adjusting the two-photon detuning of the three Raman beams to cancel externally applied acceleration and angular velocity.
While closed-loop angular velocity measurements with spatial-domain Mach–Zehnder interferometers using thermal atomic beams have been demonstrated previously~\cite{satoClosedloopMeasurementsAtominterferometer2025}, here we extend this capability to acceleration measurement by employing the digital closed-loop approach.
The stepwise nature of our method allows the frequency shift required for closed-loop acceleration measurement to be reset at the start of each cycle, preventing the unbounded accumulation of the control parameters inherent to continuous measurements.

\section{Theory}
In this section, we outline the difficulty of simultaneously measuring acceleration and angular velocity in a conventional counter-propagating thermal atomic beam interferometer, and show how this limitation is overcome by a digital closed-loop scheme.

\subsection{Conventional Counter-Propagating Atomic Beam Interferometer}
\begin{figure}[htb]
    \centering
    \includegraphics[scale=0.28]{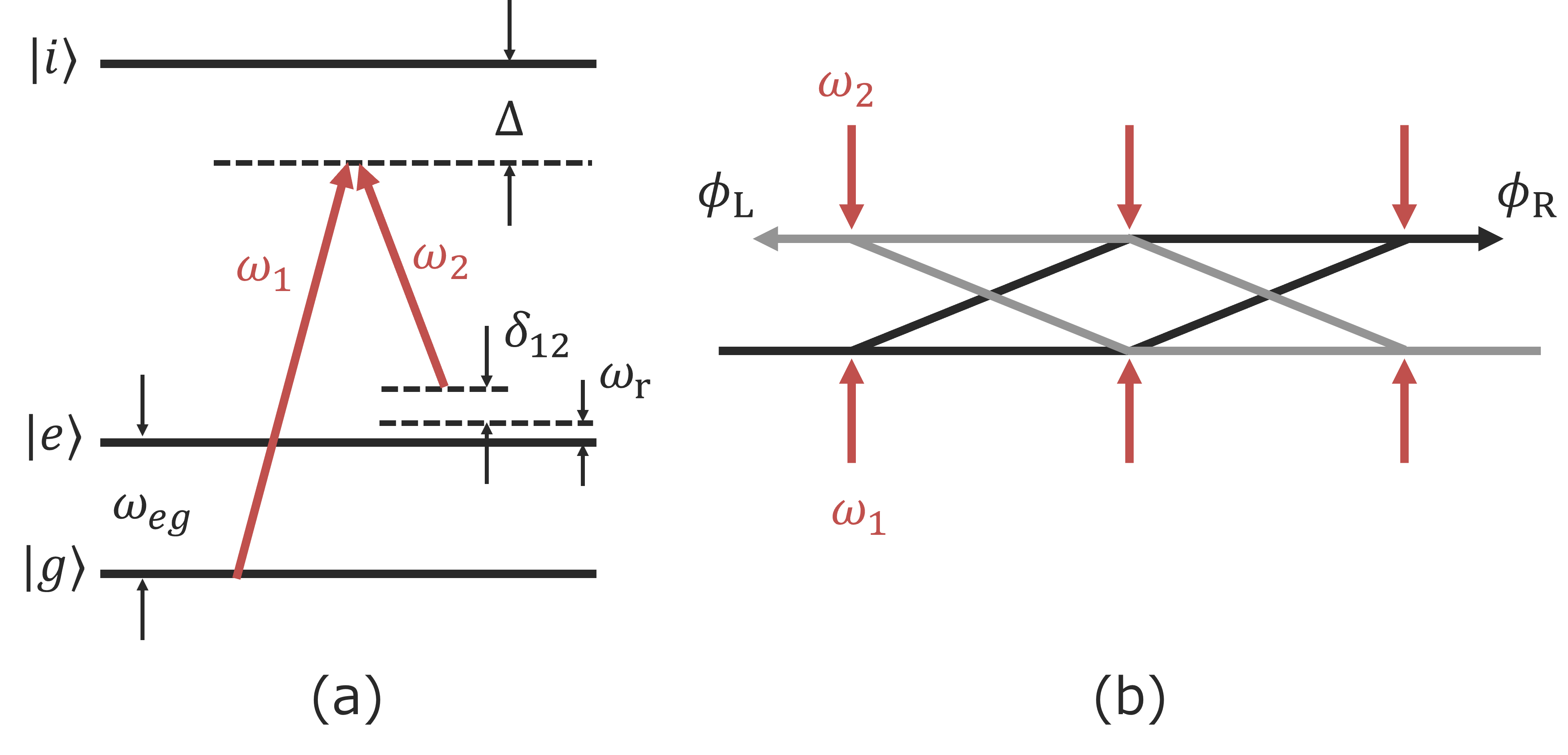}
    \caption{Atom interferometer with counter-propagating beams: (a) Level diagram of stimulated Raman transitions; (b) interferometer configuration.}
    \label{fig_counter_diagram}
\end{figure}

Figure~\ref{fig_counter_diagram}(a) depicts the energy diagram of a two-photon Raman process between the ground-state hyperfine levels $\ket{g}$ and $\ket{e}$ of an alkali atom.
The hyperfine splitting is $\omega_{eg}$, the photon recoil angular frequency is $\omega_{\rm{r}}$, and the two laser angular frequencies are $\omega_1$ and $\omega_2$.
We denote the corresponding wave 
vectors as $\boldsymbol{k}_1$ and $\boldsymbol{k}_2$.
The single-photon detuning from the excited state $\ket{i}$ is $\Delta$, and the two-photon detuning is $\delta_{12}$. For simplicity, we consider here the resonant case $\delta_{12} = 0$.
Figure~\ref{fig_counter_diagram}(b) shows a Mach–Zehnder-type quantum inertial sensor using counter-propagating atomic beams and three spatially separated Raman laser beams.
Atoms initially in $\ket{g}$ undergo Rabi oscillations to $\ket{e}$ upon interaction with the Raman beams, receiving a momentum kick and forming a superposition of two paths.
Adjusting the Raman pulse areas to $\pi/2$–$\pi$–$\pi/2$ closes the interferometer, with the resulting phase containing both inertial contributions and the relative phase between the Raman beams.

If the longitudinal velocity distribution is given by $f(v)$, the population of the $\lvert e\rangle$ state under acceleration $\boldsymbol{a}$ and angular velocity $\boldsymbol{\Omega}$ is
\begin{align}
P_e(\boldsymbol{a},\boldsymbol{\Omega}) 
&= \frac{1}{2} \int_{0}^{\infty} f(v) \times \notag \\
&\quad \biggl\{
  1 - \cos \Bigl[
    \boldsymbol{k}_{\mathrm{eff}} \cdot 
    \left( \boldsymbol{a} + 2\,\boldsymbol{\Omega} \times \boldsymbol{v} \right) 
    \left( \frac{L}{v} \right)^2 
    + \phi_{\mathrm{laser}}
  \Bigr]
\biggr\} \, dv ,
\label{Pe_integral_vec}
\end{align}
where $\boldsymbol{k}_{\mathrm{eff}} = \boldsymbol{k}_1 - \boldsymbol{k}_2$ is the effective Raman wave vector, 
and $L$ is the separation between the Raman beams.

For simplicity, consider the case where an acceleration $a$ is applied along the propagation direction of the Raman light, 
and an angular velocity $\Omega$ is applied perpendicular to the plane of the paper.  
The population of the $\lvert e\rangle$ state for the interferometer with right-oriented atomic beam in Fig.~\ref{fig_counter_diagram}(b) is then given by
\begin{equation}
\begin{aligned}
P_{e,\rm{R}}(a, \Omega) &= \frac{1}{2} \int_{0}^{\infty} f(v) \times \\
&\quad \left\{ 1 - \cos \left[ \phi_a(v) + \phi_\Omega(v) + \phi_{\mathrm{laser}} \right] \right\} \, dv
\end{aligned}
\label{Pe_integral}
\end{equation}
where  
\begin{align}
\phi_a(v) &= k_{\mathrm{eff}} \left( \frac{L}{v} \right)^2 a,
\label{S_a} \\[4pt]
\phi_\Omega(v) &= \frac{2 k_{\mathrm{eff}} L^2}{v} \, \Omega,
\label{S_omega} \\[4pt]
\phi_{\mathrm{laser}} &= \phi_{\rm{A}} - 2\phi_{\rm{B}} + \phi_{\rm{C}},
\label{S_laser}
\end{align}
and $\phi_i$ ($i = \rm{A, B, C}$) are arbitrary laser phases at each pulse.
The output phase $\phi_{\rm{R}}(v)$ is
\begin{align}
\phi_{\rm{R}}(v) &= \phi_a(v) + \phi_\Omega(v) + \phi_{\mathrm{laser}}, 
\label{phi_R}
\end{align}
Owing to the sign dependence of the velocity vector in Eq.~(\ref{Pe_integral_vec}),
the output phases $\phi_{\rm{L}}(v)$ for left-oriented atomic beam is
\begin{align}
\phi_{\rm{L}}(v) &= \phi_a(v) - \phi_\Omega(v) + \phi_{\mathrm{laser}}.
\label{phi_L}
\end{align}
From Eqs.~(\ref{phi_R}) and (\ref{phi_L}),
\begin{align}
\phi_a(v) &= \frac{\phi_{\rm{R}}(v) + \phi_{\rm{L}}(v)}{2} - \phi_{\mathrm{laser}}, 
\label{eq:phi_sum} \\
\phi_\Omega(v) &= \frac{\phi_{\rm{R}}(v) - \phi_{\rm{L}}(v)}{2}. 
\label{eq:phi_diff}
\end{align}
While $\phi_\Omega(v)$ follows directly from Eq.~(\ref{eq:phi_diff}), 
$\phi_a(v)$ in Eq.~(\ref{eq:phi_sum}) is offset by the arbitrary laser phase. 
In cold-atom experiments, identical spatial-mode Raman pulses applied at fixed intervals cancel this term. 
In thermal atomic beam interferometers, however, suppressing it is essential for determining the absolute acceleration.

\begin{figure}
    \centering
    \includegraphics[scale=0.3]{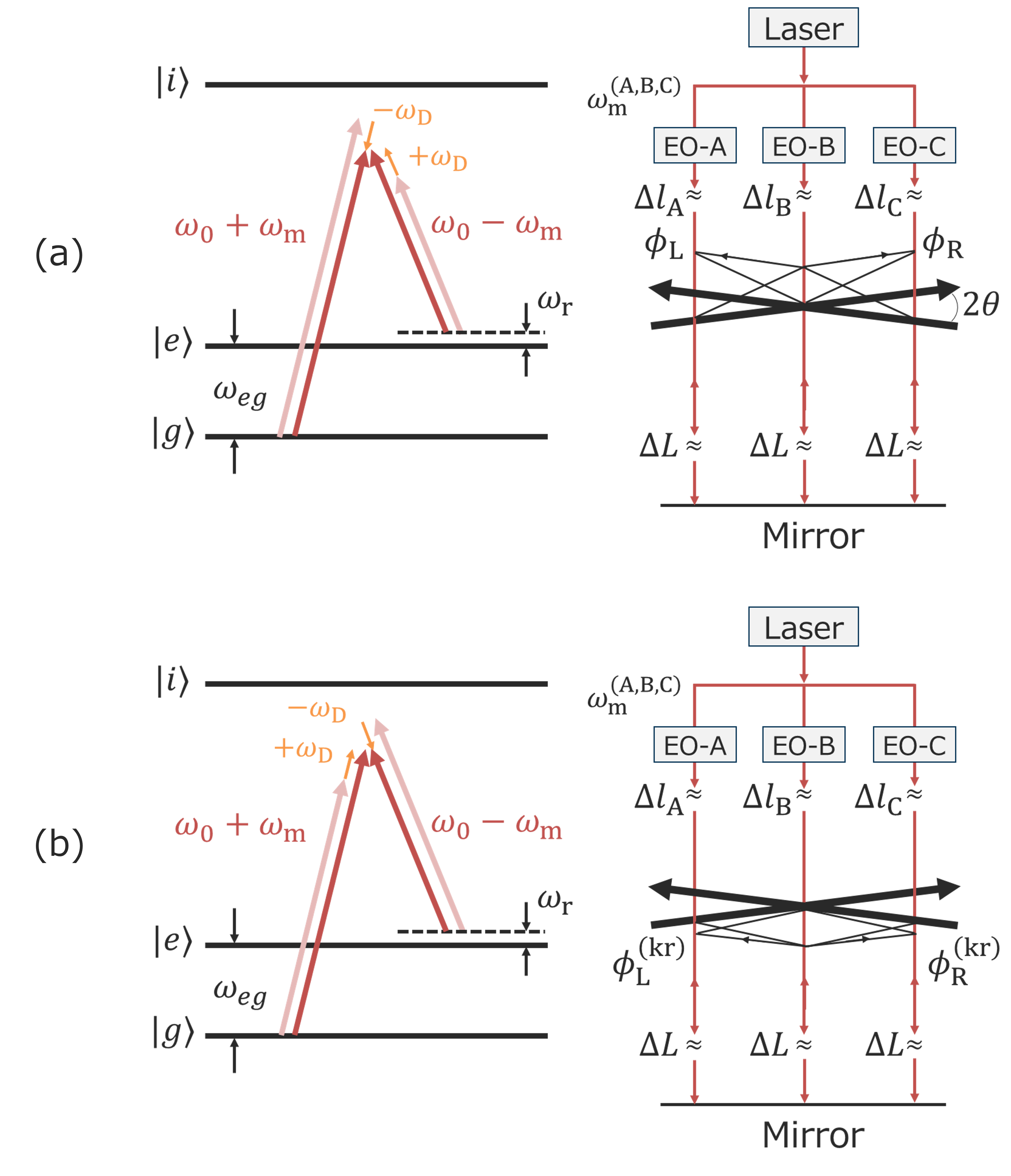}
    \caption{Counter-propagating atomic beam interferometer for digital closed-loop operation. Phase modulation at angular frequency $\omega_{\rm{m}}$ applied to the electro-optic modulator (EOM) produces sidebands at $\omega_0 \pm \omega_{\rm{m}}$ relative to the carrier angular frequency $\omega_0$. Varying $\omega_{\rm{m}}$ induces interferometer area reversal, as illustrated in (a) and (b).}
    \label{fig_mirror_scheme}
\end{figure}

\begin{figure}
    \centering
    \includegraphics[scale=0.38]{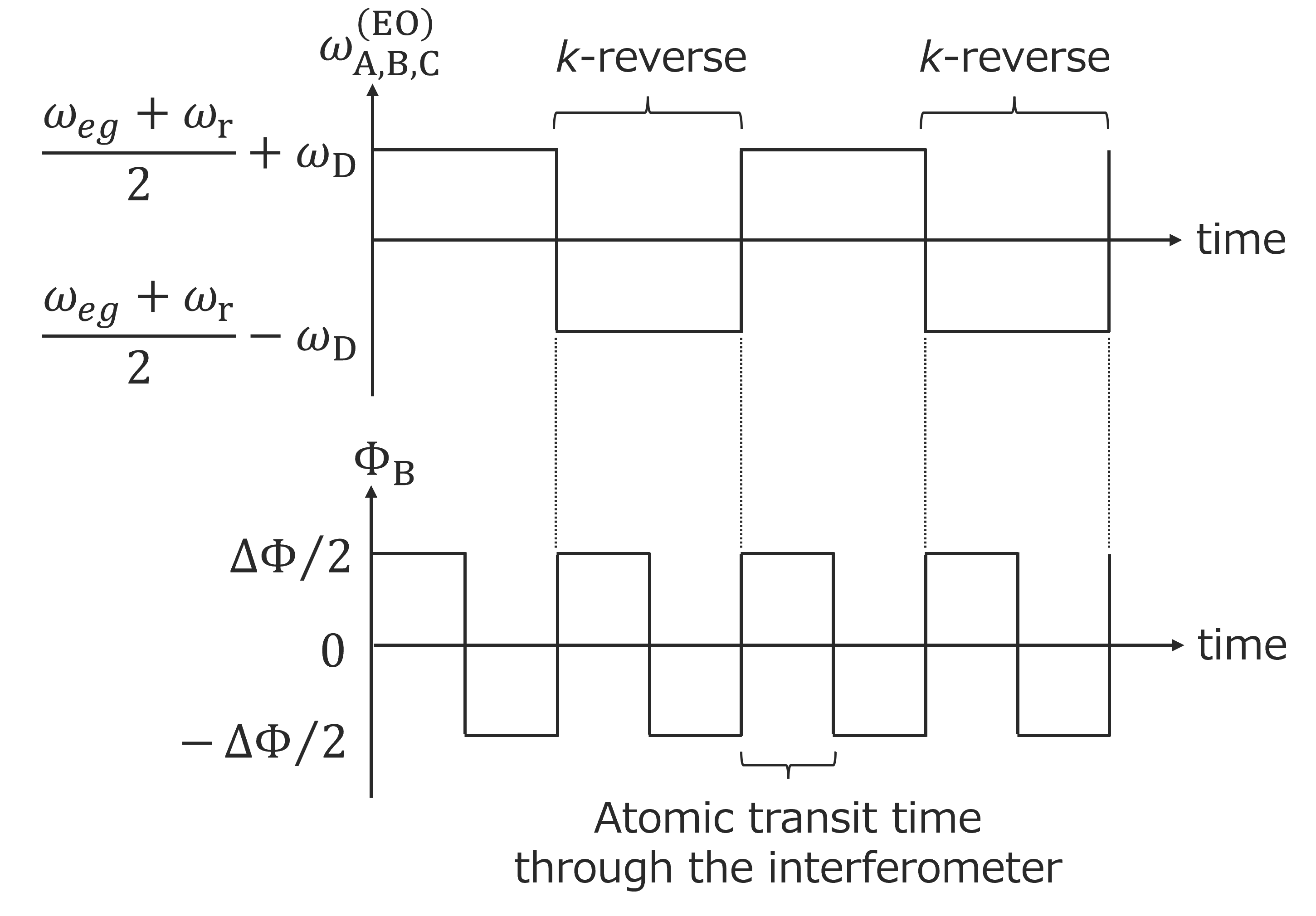}
    \caption{Timing diagram of the digital closed-loop operation. The atomic transit time at the most probable velocity $v_{\rm mp}$ through the interferometer is used as the unit. The phase $\Phi_{\rm B}$ of EO-B in Fig.~\ref{fig_mirror_scheme} is alternately biased by $\pm \Delta \Phi/2$, which, through the Raman process induced by Raman beam~B, results in a phase shift of $\pm 2 \Delta \Phi$ imparted to the atom interferometer. At twice this period, the RF frequencies driving the three EOMs are alternately shifted to perform $k$-reversal.}
    \label{fig_timing_chart}
\end{figure}

\subsection{Digital Phase-Biasing for Simultaneous Acceleration and Angular Velocity Measurement}
We extend the method proposed by Kwolek \textit{et al.}~\cite{kwolekNoiseResilienceHighBandwidth2025} and apply a digital closed-loop scheme to overcome this problem.
As shown in Fig.~\ref{fig_mirror_scheme}, phase modulation at an angular frequency $\omega_{\rm{m}}$ with respect to the laser carrier angular frequency $\omega_0$ generates two sideband components at $\omega_0+\omega_{\rm{m}}$ and $\omega_0-\omega_{\rm{m}}$.
By reflecting these laser beams with a mirror, we prepare the counter-propagating Raman beams required for constructing the interferometer.
As in Kwolek’s approach, the atomic beam is inclined by an angle $\theta$ relative to orthogonal incidence on the Raman beams.
As a modification essential for stable digital closed-loop operation, we introduce an additional counter-propagating atomic beam, also inclined by $\theta$, which does not exactly overlap with the original beam; this geometry is critical for achieving high dynamic range and simultaneous measurement of acceleration and rotation.
For an atomic beam with most probable velocity $v_{\mathrm{mp}}$, the atoms experience a Doppler shift $\omega_{\rm{D}} = k_{\mathrm{eff}} v_{\mathrm{mp}} \sin{\theta}$ relative to the Raman beams.
By varying the modulation angular frequency $\omega_{\rm{m}}$ by $\pm\omega_{\rm{D}}$, we can perform the interferometer area reversal ($k$-reversal), as illustrated in Fig.~\ref{fig_mirror_scheme}(a) and (b).
It should be noted that, unlike fast $k$-reversal schemes designed to suppress high-frequency AC noise~\cite{kwolekNoiseResilienceHighBandwidth2025}, the $k$-reversal operation in the present digital closed-loop protocol is synchronized to the atomic transit time.
Accordingly, the primary role of $k$-reversal here is to enable absolute phase reconstruction and to suppress quasi-static and slow-drift noise sources, such as optical path-length fluctuations.

Figure~\ref{fig_timing_chart} shows the timing diagram for implementing the digital closed loop.
Using the transit time of atoms with velocity $v_{\rm mp}$ through the interferometer as the unit, the phase of EOM-B is alternately biased by $\pm \Delta \Phi/2$, which, through the Raman process induced by Raman beam~B, results in a phase shift of $\pm 2 \Delta \Phi$ imparted to the atom interferometer.
Let $\phi$ denote the interferometer output phase, which includes contributions from acceleration, angular velocity, and the arbitrary laser phases.
The fluorescence intensities from atoms in state $\ket{e}$, measured at the timing of positive and negative phase shifts, are given by
\begin{align}
I_{\mathrm{up}} &= I_{\mathrm{offset}} - I_{\mathrm{amp}} \cos(\phi + 2\Delta\Phi),\\
I_{\mathrm{down}} &= I_{\mathrm{offset}} - I_{\mathrm{amp}} \cos(\phi - 2\Delta\Phi),
\end{align}
from which the interferometer output phase can be calculated.
Here, $I_{\rm{up}}$ and $I_{\rm{down}}$ denote the fluorescence intensities with the phase bias $\pm \Delta \Phi$, $I_{\rm{offset}}$ is the mean fluorescence level, and $I_{\mathrm{amp}}$ denotes the amplitude of the sinusoidal variation of the fluorescence intensity arising from the interferometric signal.
With premeasured (calibrated) $I_{\mathrm{amp}}$, the phase is obtained from
\begin{equation}
\phi = \sin^{-1}\!\left[\frac{I_{\mathrm{up}} - I_{\mathrm{down}}}{2 I_{\mathrm{amp}} \sin (2\Delta\Phi)}\right].
\label{phase_extract}
\end{equation}
In practice, two output phases, $\phi_{\rm{R}}$ and $\phi_{\rm{L}}$, are obtained from the right and left interferometers.
By alternately shifting the RF frequencies driving EOM-A, B, and C by $\pm\omega_{\rm{D}}$ with a period twice the atomic transit time, $k$-reversal is performed, yielding two additional output phases, $\phi_{\rm{R}}^{(\mathrm{kr})}$ and $\phi_{\rm{L}}^{(\mathrm{kr})}$. 
Note that precise knowledge of $I_{\mathrm{amp}}$ is unnecessary, since the closed loop maintains a quasi-inertial operating point---i.e., the system is stabilized such that all four output phases are zero---and the required feedback signal is used as the readout, as will be discussed later.

Using the phases obtained above, $\phi_{\rm{R}}$, $\phi_{\rm{L}}$, $\phi_{\rm{R}}^{(\mathrm{kr})}$, and $\phi_{\rm{L}}^{(\mathrm{kr})}$, we first cancel the arbitrary laser phase.
To do so, we explicitly evaluate the laser phase at the atom--light interaction point for each of the three Raman beams.
A carrier laser beam emitted from the light source can be represented as $E(z) = E_0 \sin(k_0 z - \omega_0 t)$, where \( E_0 \), $k_0$ and \( \omega_0 \) are the amplitude, wavenumber and angular frequency of the carrier, respectively.
EOMs generate the required sidebands for the counter-propagating Raman transitions, with wavenumbers $k_1$ and $k_2$.
Atoms entering the interferometer interact with the downward-propagating Raman beam at positions \(z_i = l_i + \Delta l_i\) for \(i = \rm{A,B,C}\),
where \(l_i\) denotes the optical path length from EOM-\(i\) to the nominal atom position in the absence of momentum kicks, and \(\Delta l_i\) represents its temporal fluctuation. Note that the displacement of the atomic trajectory due to the momentum kicks is accounted for in the Sagnac phase shift and can therefore be neglected here.
The optical path length from the interaction point to the mirror is \(L_i\), and \(\Delta L\) represents the temporal variation of the mirror position.
In the interferometer shown in Fig.~\ref{fig_mirror_scheme}(a), the Raman optical phase imprinted on the atoms, \(\phi_i\), is, for either of the counter-propagating atomic beams, given by
\begin{equation}
\phi_i = (k_1 - k_2)(l_i + \Delta l_i) + 2k_1(L_i + \Delta L).    
\end{equation}

The total accumulated phase for the interferometer can be expressed as
\begin{align}
\phi_{\mathrm{laser}}
&= \phi_{\rm{A}} - 2\phi_{\rm{B}} + \phi_{\rm{C}} \notag\\
&= (k_1 - k_2)\lbrack(\Delta l_{\rm{A}} - 2\Delta l_{\rm{B}} + \Delta l_{\rm{C}}) \notag\\
&\quad + (l_{\rm{A}} - 2l_{\rm{B}} + l_{\rm{C}})\rbrack.
\end{align}

Similarly, for the interferometer shown in Fig.~\ref{fig_mirror_scheme}(b), the laser-phase contribution to the interferometer phase is given by
\begin{align}
\phi_{\mathrm{laser}}^{(\mathrm{kr})}
&= \bigl( k_1^{(\mathrm{kr})} - k_2^{(\mathrm{kr})}\bigr)
   \lbrack(\Delta l_{\rm{A}} - 2\Delta l_{\rm{B}} + \Delta l_{\rm{C}}) \notag\\
&\quad + (l_{\rm{A}} - 2l_{\rm{B}} + l_{\rm{C}})\rbrack.
\end{align}
From $\phi_\mathrm{laser}$ and $\phi^\mathrm{(kr)}_\mathrm{laser}$, it follows that fluctuations in the mirror position are not imprinted onto the interferometer phase.
Moreover, even if the distance from the source to the atomic beam fluctuates, the wavenumber governing the phase sensitivity is $k_1 - k_2$ rather than $k_{\mathrm{eff}}$.
Consequently, the sensitivity to such path-length fluctuations
is suppressed by roughly five orders of magnitude compared with a configuration employing two counter-propagating beams delivered through independent optical paths.

Here, based on the above results, we examine the four interferometer phases corresponding to atoms with velocity $v$.
\begin{align}
\Lambda &\equiv (\Delta l_{\rm{A}} - 2\Delta l_{\rm{B}} + \Delta l_{\rm{C}}) + (l_{\rm{A}} - 2l_{\rm{B}} + l_{\rm{C}}), \label{eq:Lambda}
\end{align}

\begin{align}
\phi_{\rm{R}}(v) &= \phi_a(v) + \phi_\Omega(v) + (k_1 - k_2)\Lambda,\\
\phi_{\rm{R}}^{(\mathrm{kr})}(v) &= -\phi_a(v) - \phi_\Omega(v)
  + \bigl(k_1^{(\mathrm{kr})} - k_2^{(\mathrm{kr})}\bigr)\Lambda,\\
\phi_{\rm{L}}(v) &= \phi_a(v) - \phi_\Omega(v) + (k_1 - k_2)\Lambda,\\
\phi_{\rm{L}}^{(\mathrm{kr})}(v) &= -\phi_a(v) + \phi_\Omega(v)
  + \bigl(k_1^{(\mathrm{kr})} - k_2^{(\mathrm{kr})}\bigr)\Lambda.
\end{align}

The parameter $\Lambda$—which captures the optical path-length differences and temporal fluctuations of the three Raman beams—can be written as a combination of the four interferometer phases:
\begin{equation}
\Lambda = \frac{\bigl[\phi_{\rm{R}}(v)+\phi_{\rm{R}}^{(\mathrm{kr})}(v)\bigr] + \bigl[\phi_{\rm{L}}(v)+\phi_{\rm{L}}^{(\mathrm{kr})}(v)\bigr]}{8\,\omega_{\rm{m}}}c,
\end{equation}
where $c$ is the speed of light in the vacuum.

As seen from Eq.~(\ref{Pe_integral}), it is not possible to extract the interference signal arising from atoms with a specific velocity $v$ individually. 
What we can observe is always the interference signals $\phi_{\rm{R}}$, $\phi_{\rm{R}}^{(\mathrm{kr})}$, $\phi_{\rm{L}}$, and $\phi_{\rm{L}}^{(\mathrm{kr})}$, each of which contains contributions from the entire velocity distribution. 
However, by applying feedback to the optical system such that
$\phi_{\rm{R}}+\phi_{\rm{R}}^{(\mathrm{kr})}+\phi_{\rm{L}}+\phi_{\rm{L}}^{(\mathrm{kr})}$
is set to zero, it is possible to make $\Lambda$ vanish. 

The parameter $\Lambda$ is primarily influenced by temperature-induced drift once the path length difference has been sufficiently minimized in advance—through fiber length adjustments or mechanical jigs. Even with several meters of optical fiber, if the overall system temperature is stabilized within approximately \SI{1}{K}, $\Lambda$ remains on the order of \SI{100}{\micro m}.
As discussed earlier, because the wavenumber governing phase sensitivity is $k_1 - k_2$ rather than $k_\mathrm{eff}$, the resulting effect on the interferometer’s output phase is less than \SI{1}{\degree}.
Moreover, slow feedback, for example by adjusting the optical path length using an optical phase shifter, can further reduce $\Lambda$ to negligible levels.
In the digital closed loop, with feedback applied so that $\Lambda$ is nulled for every phase-biasing/$k$-reversal set, the interferometer phases associated with acceleration and angular velocity are given by
\begin{align}
\phi_a(v) &= \frac{\bigl[\phi_{\rm{R}}(v)-\phi_{\rm{R}}^{(\mathrm{kr})}(v)\bigr]+\bigl[\phi_{\rm{L}}(v)-\phi_{\rm{L}}^{(\mathrm{kr})}(v)\bigr]}{4},
\label{phi_a_open}\\
\phi_\Omega(v) &= \frac{\bigl[\phi_{\rm{R}}(v)-\phi_{\rm{R}}^{(\mathrm{kr})}(v)\bigr]-\bigl[\phi_{\rm{L}}(v)-\phi_{\rm{L}}^{(\mathrm{kr})}(v)\bigr]}{4}.
\label{phi_omega_open}
\end{align}

Note that in the present theoretical model, only the first-order modulation sidebands are considered.
In practice, phase modulation by an EOM generates higher-order sidebands, which may give rise to parasitic interferometers, as discussed in Ref.~\cite{Carraz2012}.
Such effects are not included in the present analysis but can be effectively suppressed in experiments by optical filtering or by employing single-sideband EOMs that eliminate unwanted sidebands~\cite{Zhu2018}. A detailed analysis of these technical effects is left for future experimental work.

\subsection{Closed-Loop Acceleration and Angular Velocity Measurement}
Given that the influence of arbitrary laser phase is eliminated, Eqs.~(\ref{S_a}) and (\ref{S_omega}) can be used to extract acceleration and angular velocity by representing $v$ with $v_{\mathrm{mp}}$ as
\begin{align}
a &\approx 
\frac{\bigl[\phi_{\rm{R}}-\phi_{\rm{R}}^{(\mathrm{kr})}\bigr]+\bigl[\phi_{\rm{L}}-\phi_{\rm{L}}^{(\mathrm{kr})}\bigr]}
     {4 k_{\mathrm{eff}}} \left( \frac{v_{\mathrm{mp}}}{L} \right)^2, \\[4pt]
\Omega &\approx  \frac{\bigl[\phi_{\rm{R}}-\phi_{\rm{R}}^{(\mathrm{kr})}\bigr]-\bigl[\phi_{\rm{L}}-\phi_{\rm{L}}^{(\mathrm{kr})}\bigr]}{8 k_{\mathrm{eff}}}  \frac{v_{\mathrm{mp}}}{L^2}.
\end{align}

However, such open-loop measurements are accompanied by several issues. In principle, the above equations can be corrected by accurately accounting for the velocity distribution in advance.
However, even with such corrections, the sensitivity to the oven temperature remains.
They are also affected by slight variations in the atomic beam velocity caused by motion of the measurement apparatus. 
Furthermore, at higher accelerations and angular velocities, the velocity distribution of the thermal atomic beam degrades the interference contrast, thereby reducing the sensitivity of the interferometer. 
This effect not only limits the dynamic range but also, as indicated by Eq.~(\ref{phase_extract}), introduces phase measurement errors through variations in $I_\mathrm{amp}$, ultimately degrading the overall measurement accuracy of the interferometer. 
To mitigate these effects, the interferometer should be operated in a closed-loop configuration that maintains it effectively in a quasi-inertial frame at all times.

In an atom interferometer, the phase shift induced by acceleration or angular velocity arises from changes in the atomic position and velocity with respect to the traveling standing wave formed by the Raman beams.
Therefore, in order to implement a closed loop, it is important to track the two-photon detuning of the Raman beams at the atomic position, as expressed by the following equation~\cite{youngPrecisionAtomInterferometry1997}:

\begin{equation}
\delta_{12} = (\omega_1 - \omega_2) - \left( \omega_{eg} + \frac{\boldsymbol{p} \cdot \boldsymbol{k}_{\mathrm{eff}}}{m} + \omega_{\rm{r}} \right),
\end{equation}
where $\boldsymbol{p}$ is the atomic momentum vector and $m$ is the mass of atom.
First, in Fig.~\ref{fig_mirror_scheme}, we consider atoms with the most probable velocity $v_{\rm{mp}}$ within the velocity distribution, for which the two-photon resonance condition $\delta_{12} = 0$ is satisfied, assuming that no acceleration or angular velocity is applied.
If we denote the deviation from the most probable atomic velocity by $\Delta v$, this effect appears as $\Delta v \sin \theta$.
Let $v_\Omega$ and $v_a$ be the effective velocities of the Raman phase front
with respect to the atomic beam due to angular velocity and acceleration, respectively.
The two-photon detunings for Raman beams A, B, and C are then given by

\begin{align}
\delta_{12}^{(\rm{A})} &= -k_{\mathrm{eff}} \left( v_\Omega^{(\rm{A})} + v_a^{(\rm{A})} + \Delta v \sin \theta \right), \\
\delta_{12}^{(\rm{B})} &= -k_{\mathrm{eff}} \left( v_a^{(\rm{B})} + \Delta v \sin \theta \right), \\
\delta_{12}^{(\rm{C})} &= -k_{\mathrm{eff}} \left( v_\Omega^{(\rm{C})} + v_a^{(\rm{C})} + \Delta v \sin \theta \right).
\end{align}

Here, we set the two-photon detuning for Raman beam A to $\delta + \gamma t$, for beam B to $\delta$, and for beam C to $\delta - \gamma t$.  
Let $T_{\mathrm{oven}}$ be the time from
the atom leaving the oven to its reaching Raman beam A
and $T$ the time taken to traverse the interferometer arm length $L$ (see Fig.~\ref{fig_interference_calc}). 
It is evident that $T_{\mathrm{oven}}$ and $T$ are functions of the atomic velocity $v$.
Then, expressing $v_\Omega$ and $v_a$ explicitly, we have

\begin{align}
\delta_{12}^{(\rm{A})} &= \delta + \gamma T_{\mathrm{oven}}
 - k_{\mathrm{eff}}\!\left( \Omega L + a T_{\mathrm{oven}} + \Delta v \sin \theta \right), 
\label{delta12A}\\
\delta_{12}^{(\rm{B})} &= \gamma (T_{\mathrm{oven}} + T)
 - k_{\mathrm{eff}}\!\left[ a (T_{\mathrm{oven}} + T) + \Delta v \sin \theta \right], \\
\delta_{12}^{(\rm{C})} &= -\delta + \gamma (T_{\mathrm{oven}} + 2T) \notag\\
&\quad - k_{\mathrm{eff}}\!\left[-\Omega L + a (T_{\mathrm{oven}} + 2T) + \Delta v \sin \theta \right].
\end{align}

From these expressions, it is evident that by setting
\begin{align}
\delta = k_{\mathrm{eff}} \, \Omega L, \\
\gamma = k_{\mathrm{eff}} \, a,
\end{align}
eliminates the effects of acceleration and angular velocity for atoms of arbitrary velocity. 
As noted above, $\phi_a(v)$ and $\phi_\Omega(v)$ cannot be measured directly. 
Instead, $\delta$ and $\gamma$ can be obtained by adjusting them so that the following two phases vanish:
\begin{align}
\phi_a &= \frac{\bigl[\phi_{\rm{R}}-\phi_{\rm{R}}^{(\mathrm{kr})}\bigr]+\bigl[\phi_{\rm{L}}-\phi_{\rm{L}}^{(\mathrm{kr})}\bigr]}{4},
\label{phi_a}\\
\phi_\Omega &= \frac{\bigl[\phi_{\rm{R}}-\phi_{\rm{R}}^{(\mathrm{kr})}\bigr]-\bigl[\phi_{\rm{L}}-\phi_{\rm{L}}^{(\mathrm{kr})}\bigr]}{4}.
\label{phi_omega}
\end{align}

\begin{figure}
    \centering
    \includegraphics[scale=0.3]{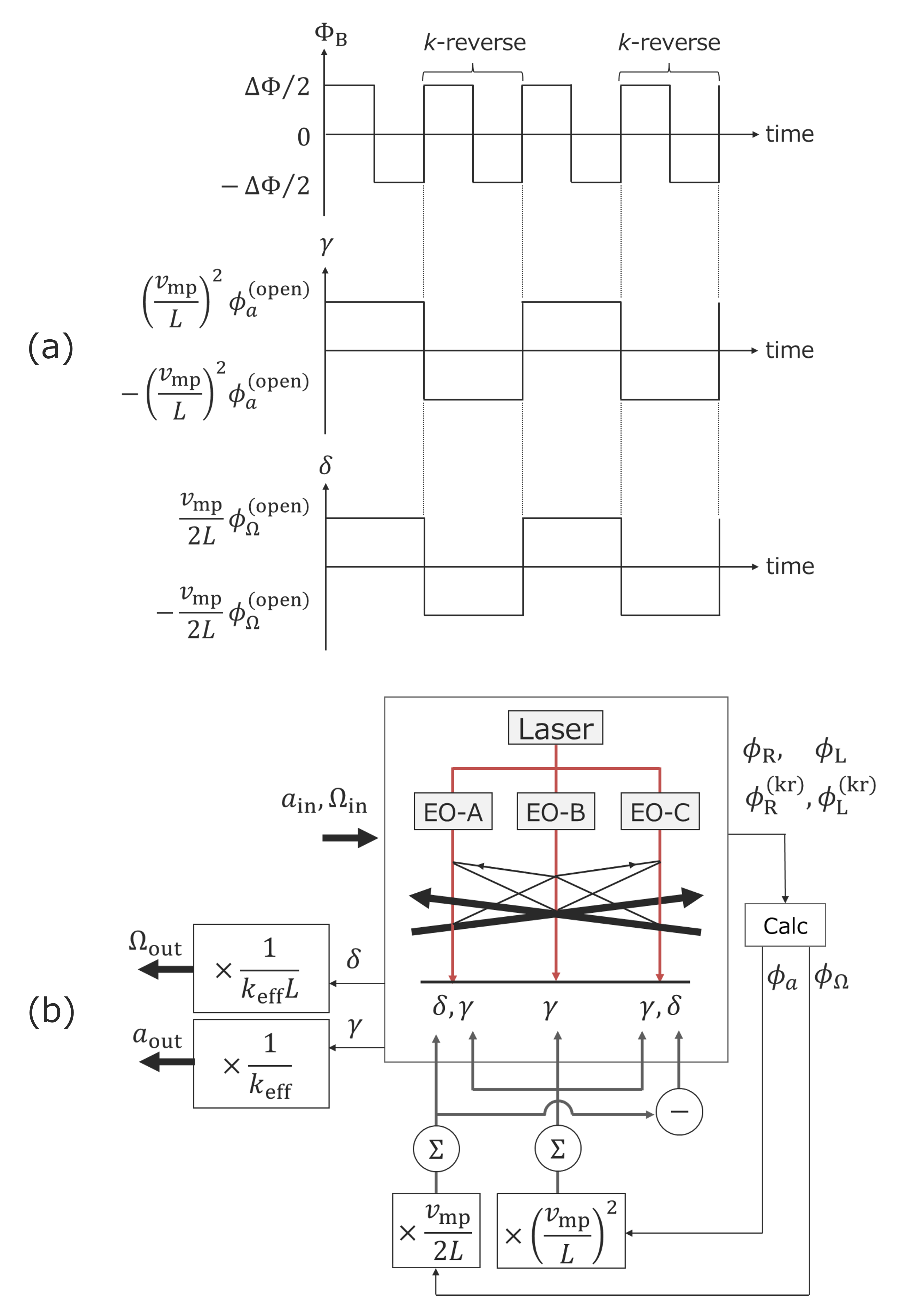}
    \caption{Digital closed loop utilizing the four phase outputs of the interferometer. 
(a) Timing chart of the two-photon detuning of the Raman light under constant angular velocity and acceleration. 
(b) Block diagram of the digital closed loop under time-varying acceleration and angular velocity. 
Here, $\phi_a^{\mathrm{(open)}}$ and $\phi_\Omega^{\mathrm{(open)}}$ denote the values of $\phi_a$ and $\phi_\Omega$ obtained in the open-loop configuration under constant acceleration and angular velocity.}
    \label{fig_DCL}
\end{figure}

Figure~\ref{fig_DCL}(a) shows the time chart of the two-photon Raman detunings, $\delta$ and $\gamma$, which are required to maintain the interferometer in a quasi-inertial frame under constant acceleration and angular velocity. 
When the $k$-reversal operation is performed, the direction of the recoil momentum imparted to the atoms is reversed, and therefore the signs of $\delta$ and $\gamma$ must also be inverted. 
Simultaneously, resetting the two-photon detuning suppresses the temporal drift caused by $\gamma$. 
The amplitudes of $\delta$ and $\gamma$ can be approximately estimated from Eqs.~(\ref{S_a}) and (\ref{S_omega}) as
\begin{equation}
\delta \approx \frac{v_{\mathrm{mp}}}{2L} \, \phi_\Omega^{({\rm open})},
\end{equation}
\begin{equation}
\gamma \approx \left( \frac{v_{\mathrm{mp}}}{L} \right)^{2} \phi_a^{({\rm open})}.
\end{equation}
Here, $\phi_a^{\mathrm{(open)}}$ and $\phi_\Omega^{\mathrm{(open)}}$ denote the values of $\phi_a$ and $\phi_\Omega$ obtained in the open-loop configuration under constant acceleration and angular velocity.

Figure~\ref{fig_DCL}(b) presents a block diagram of the digital closed loop in a case where the applied acceleration and angular velocity vary in real time.
As will be detailed in the simulations, the required changes in two-photon detuning to cancel the variations in external acceleration and angular velocity are computed and fed back to the system.
As a result, the system response is effectively confined within four times the atomic transit time through the interferometer, enabling a high bandwidth.
Moreover, because the interferometer is maintained in a quasi-inertial frame at all times, no loss of interference contrast occurs due to the atomic velocity distribution, ensuring a high dynamic range.
Acceleration and angular velocity are derived from the two-photon detuning characterized by $\delta$ and $\gamma$, respectively.
This enables simultaneous measurement and ensures high scale-factor stability of the sensor, as it is unaffected by fluctuations of atomic velocity.\par

\section{Simulation}
In this section, we perform numerical simulations of a digital closed-loop quantum inertial sensor using $^{85}\mathrm{Rb}$ atoms emitted from an oven, in order to verify the validity of the theoretical proposal and to estimate the achievable sensitivity. 

Since the method of Kwolek \textit{et al.}~\cite{kwolekNoiseResilienceHighBandwidth2025} was developed for cold atomic beams, it is first necessary to confirm whether an interferometer employing retroreflection mirrors can also operate with a thermal atomic beam.  
In this context, the narrow transverse momentum spread of the atomic beam is fundamentally important.  
Therefore, we assume the use of a commercially available capillary plate (Hamamatsu Photonics, Ltd.) with an effective diameter of 6.5 mm, a thickness of 1 mm, an aspect ratio of 1000, and an open-area ratio of 30\%, employed as the atomic oven.
Atoms transmitting without wall collisions emerge within a narrow solid angle determined by the aspect ratio.  
For simplicity, wall-reflected atoms---which have a much larger divergence and negligible contribution to the interference---are excluded under a ``dark-wall'' condition.  
A source temperature of $170\,^\circ\mathrm{C}$, comparable to that used in previous cesium experiments~\cite{gustavsonPrecisionRotationSensing2000,gustavsonPrecisionRotationMeasurements1997a,gustavsonRotationSensingDual2000}, is assumed, yielding an atomic flux of \SI{7.8E10}{atoms/s}~\cite{ramseyMolecularBeams1986}.
The interferometer arm length is set to $L = 100\,\mathrm{mm}$ and the atomic beam inclination to $\theta = 0.2^\circ$, with the divergence of atoms passing through the capillary plate being smaller than $\theta$.  
The Raman beams experience a Doppler shift difference of $2\omega_{\rm{D}} = 2\pi\times 5.3\,\mathrm{MHz}$ between the area-reversed and non-reversed configurations.  
A Gaussian fit to the Maxwell--Boltzmann distribution centered at $v_{\rm{mp}}$ yields a standard deviation of $\sigma_v = 140\,\mathrm{m/s}$, corresponding to a Doppler width of  
$2k_{\mathrm{eff}}\sigma_v\sin\theta = 2\pi\times 1.3\,\mathrm{MHz}.$
The transverse velocity determined by the capillary plate's aspect ratio is $v_t \approx 0.29\,\mathrm{m/s}$, giving a corresponding Doppler width of  
$k_{\mathrm{eff}} v_t = 2\pi\times 0.75\,\mathrm{MHz}.$  
The total Doppler width is thus $1.5\,\mathrm{MHz}$.  
Under these conditions, the fraction of atoms in the selected interferometer that contribute to the opposite (area-reversed) interferometer is on the order of $2\times10^{-3}$, confirming the feasibility of implementing the proposed protocol.

The digital closed-loop simulation was conducted using the analytical expressions for the two-photon Raman process derived by B. Young \textit{et al.}~\cite{youngPrecisionAtomInterferometry1997}.
In this calculation, not only the closed interferometer paths but also the non-interfering components are included, enabling an accurate determination of the interferometric contrast.
For example, for an atom interferometer propagating to the right, the two-photon Raman process must be evaluated 12 times.
In the present simulation, both left- and right-propagating paths as well as contributions from area reversal are considered, resulting in a total of 48 evaluations.
A detailed description of this procedure is beyond the scope of the present paper and is therefore omitted. The main equations used and an example calculation are provided in the Appendix.

Figure~\ref{fig_interference} shows the simulated interference signal obtained when sweeping the phase of Raman beam~A, one of the three Raman beams constituting the interferometer. A laser beam diamiter is adjusted such that $\pi/2$--$\pi$--$\pi/2$ pulses are formed for the most probable velocity $v_{\rm{mp}} = 294\,\mathrm{m/s}$. As noted earlier, although the velocity distribution is broad with $\sigma_v = 140\,\mathrm{m/s}$, a high contrast of 52\% is obtained in the stationary state.

\begin{figure}
    \centering
    \includegraphics[scale=0.4]{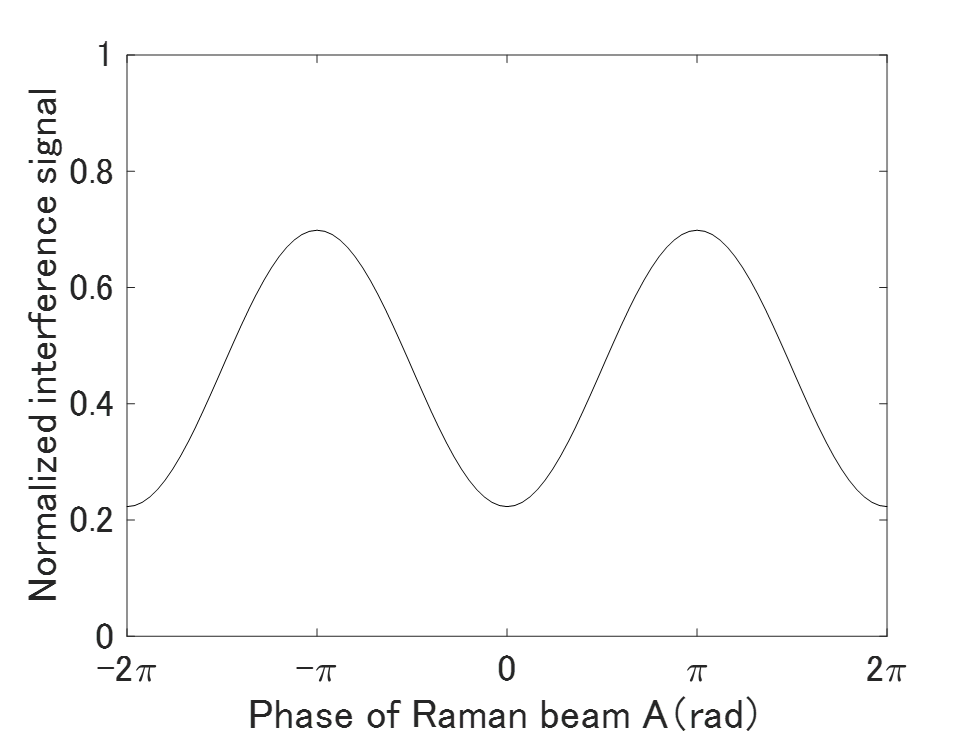}
    \caption{Simulated interference signal for an oven temperature of 
$170\,^\circ\mathrm{C}$ and an atomic beam inclination of $\theta = 0.2^\circ$.\\
The horizontal axis denotes the phase of Raman beam~A among the three beams forming the interferometer.}
    \label{fig_interference}
\end{figure}

\begin{figure}
    \centering
    \includegraphics[scale=0.3]{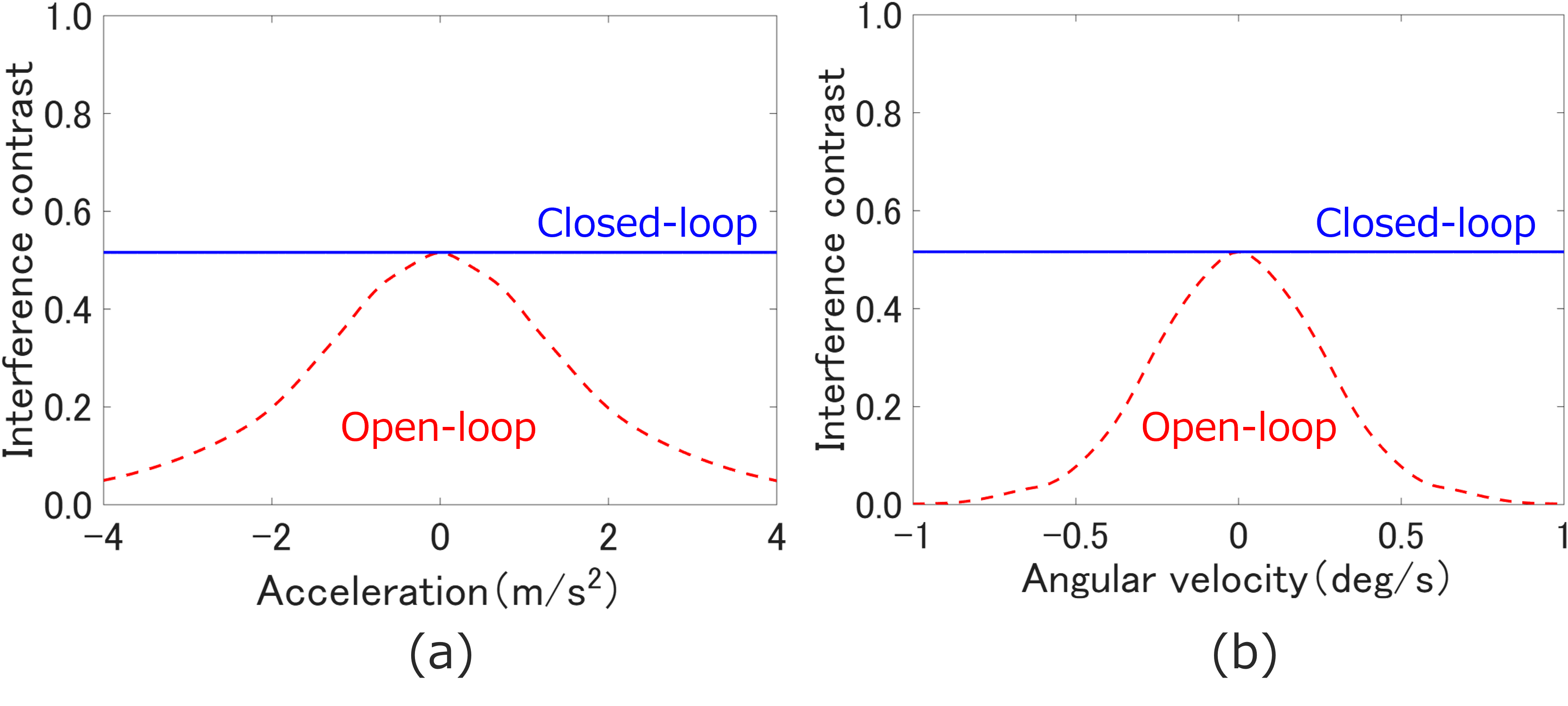}
\caption{Simulated dependence of the interferometer contrast on acceleration and angular velocity. Dashed lines indicate the open-loop case, and solid lines indicate the closed-loop case. Dependence of the interferometer contrast on (a) acceleration and (b) angular velocity.}
    \label{fig_dynamic_range}
\end{figure}

Figure~\ref{fig_dynamic_range} plots the dependence of the interference contrast on the externally applied (a) acceleration and (b) angular velocity.
The dashed curves correspond to the open-loop case, where acceleration and angular velocity are derived from the output phase of the interferometer without feedback to the two-photon detuning of the Raman beams.
The solid curves correspond to the closed-loop case, where acceleration and angular velocity are determined from the two-photon detuning under feedback control. These results indicate that digital closed-loop processing effectively avoids the contrast degradation arising from the velocity spread of the thermal atomic beam for both acceleration and angular velocity, thereby achieving a wide dynamic range.

\begin{figure}
    \centering
    \includegraphics[scale=0.4]{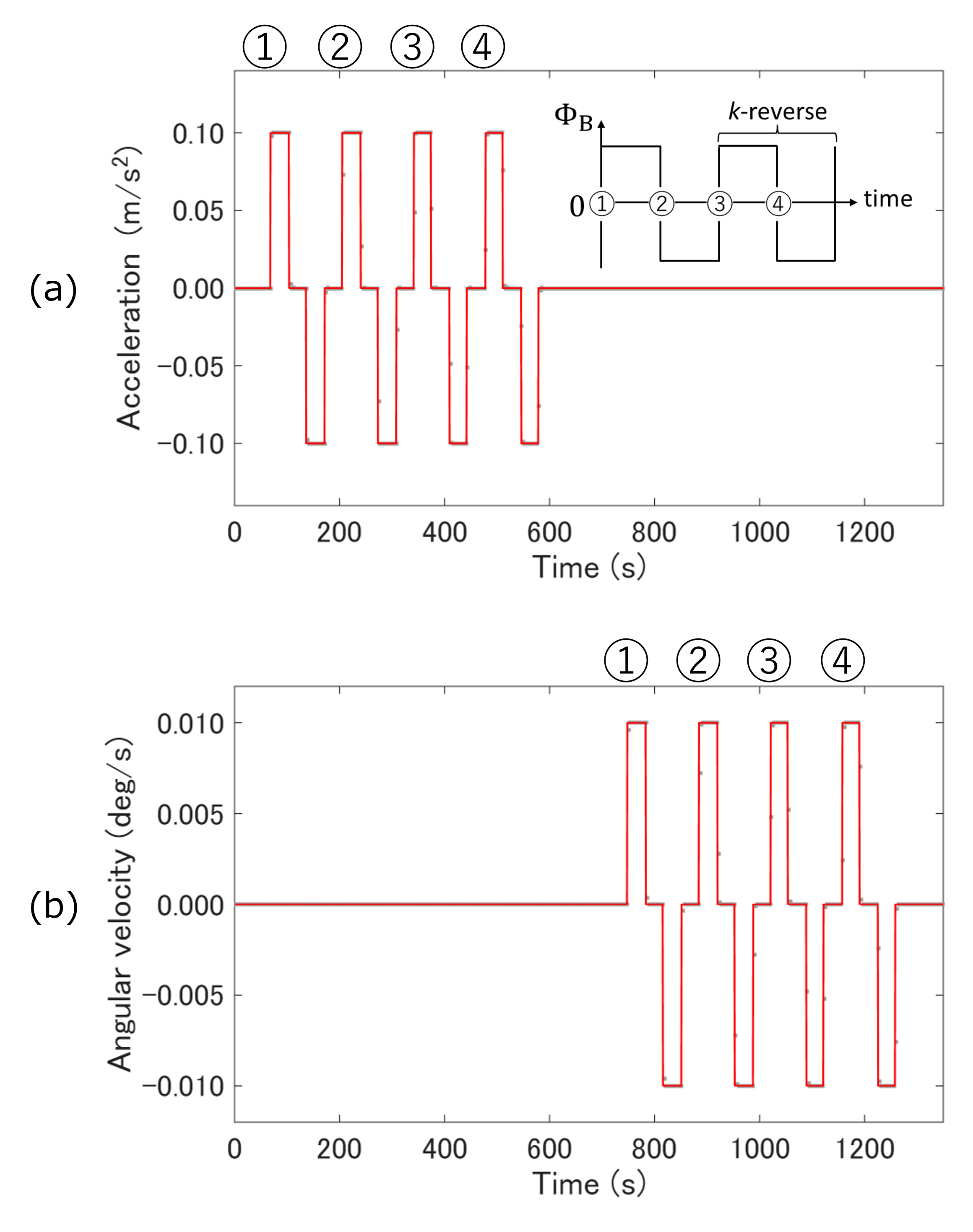}
    \caption{
    Step response of the digital closed-loop method to (a) acceleration and (b) angular velocity. The input acceleration and angular velocity are shown as red lines, while the corresponding interferometer outputs are shown as gray dots. The inset illustrates the timing at which the acceleration and angular velocity are stepped during the phase modulation of Raman beam B and the k-reversal process (cases \ding{192}–\ding{195}). Since the interferometer output closely matches the input signals, the gray output markers are almost entirely overlapped with the red input lines. The apparent scatter in the output does not represent measurement noise but arises from deterministic transient responses that depend on the relative timing between the step input and the measurement cycle.
    } 
    \label{fig_step_response}
\end{figure}

Figure~\ref{fig_step_response} shows simulated digital closed-loop responses to step changes in acceleration and angular velocity applied at different times. The input acceleration and angular velocity are shown as red lines, while the corresponding interferometer outputs are shown as gray dots. Since the interferometer output closely matches the input signals, the gray output markers are almost entirely overlapped with the red input lines for most of the time in the figures.
As illustrated in the inset, in the digital closed-loop operation, the four interferometer phases required for acceleration and angular velocity sensing are not obtained simultaneously, but rather through a sequence of four processes: alternating phase biasing of Raman beam~B and $k$-reversal.  
When step-like acceleration and angular velocity are applied externally, the onset timing is, of course, arbitrary; these cases are labeled \ding{192}--\ding{195}.  
Case~\ding{192} represents the most favorable situation, in which the four required processes are executed after both acceleration and angular velocity have stepped up.  
In this case, the calculated feedback value is accurate, and the output responds without delay in the minimum time $(2L / v_{\rm{mp}} \times 4 = 2.7~\mathrm{ms})$.
In all other cases, acceleration or angular velocity steps occur during the acquisition of the interferometer phases; nevertheless, the output responds in about $\SI{5}{\milli\second}$, ensuring a bandwidth sufficient for inertial navigation.  It is particularly noteworthy that, in all cases \ding{192}--\ding{195}, the applied acceleration and angular velocity do not couple into each other's output---i.e., no cross-coupling occurs.  
While atomic interferometers are intrinsically sensitive to both acceleration and angular velocity, such dual sensitivity can hinder accurate estimation when one effect must be distinguished from the other.
The results here indicate that this drawback can be eliminated by employing the digital closed loop, which enables simultaneous measurement of both quantities.

\begin{figure}
    \centering
    \includegraphics[scale=0.4]{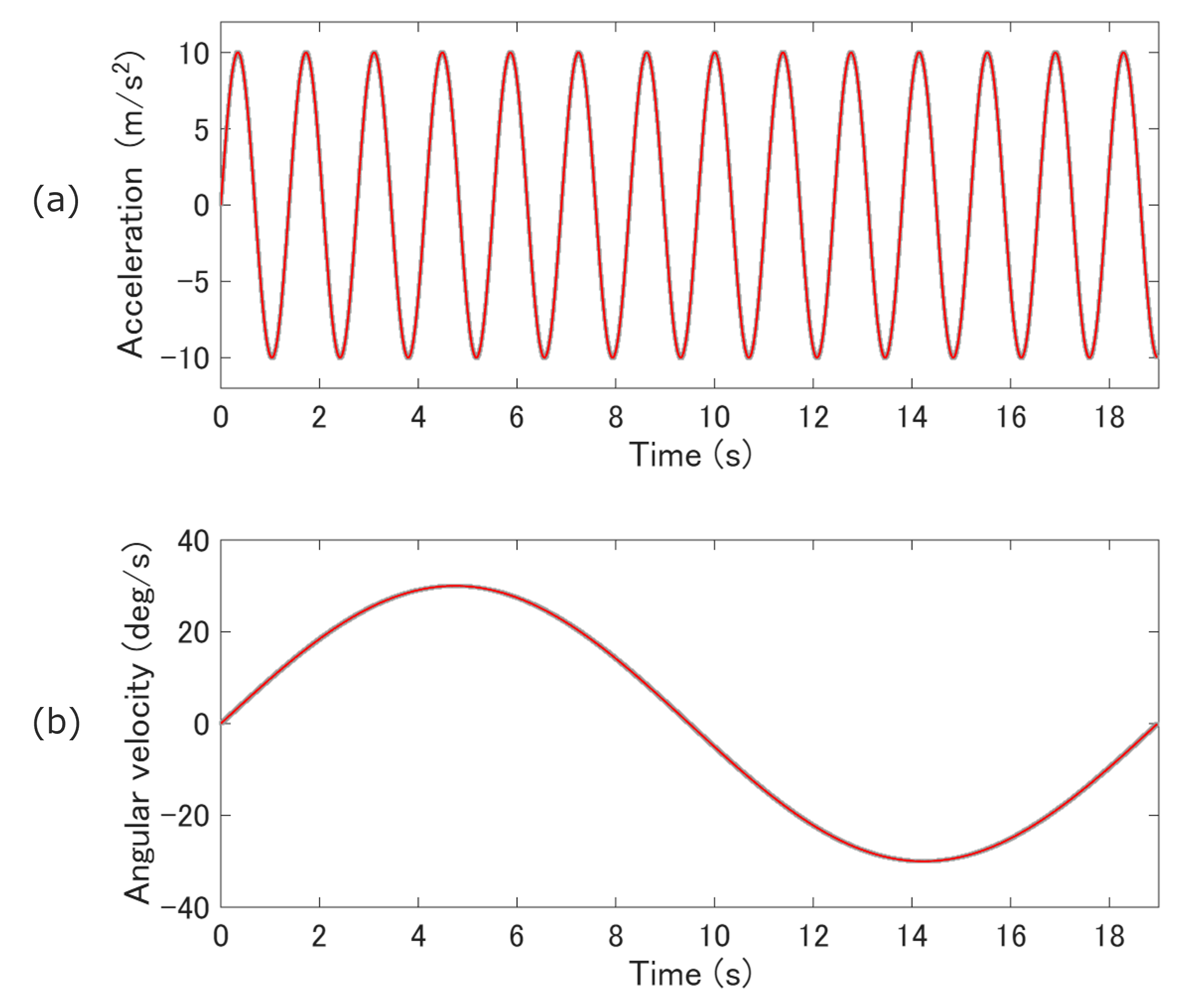}
    \caption{
    Simulation results for sinusoidally varying (a) acceleration and (b) angular velocity. In the simulation, the acceleration and angular velocity are varied simultaneously, and the two panels show the corresponding closed-loop responses, including both the input signals and the interferometer outputs. The input acceleration and angular velocity are shown as red lines, while the corresponding interferometer outputs are shown as gray dots. Since the interferometer output closely matches the input signals, the gray output markers are almost entirely overlapped with the red input lines.
}
    \label{fig_sinusoidal_response}
\end{figure}

Figure~\ref{fig_sinusoidal_response} shows the simulated response of the digital closed loop when externally applied acceleration and angular velocity vary sinusoidally, including both the applied input signals and the corresponding interferometer outputs. The input acceleration and angular velocity are plotted as red lines, while the interferometer outputs are shown as gray markers. Since the interferometer output closely matches the input signals, the gray output markers are almost entirely overlapped with the red input lines.
Even when an acceleration equivalent to gravitational acceleration is applied with alternating sign, together with a large angular velocity on the order of several tens of degrees per second, the system responds properly. The simulation includes the longitudinal velocity distribution of the thermal atomic beam and the digital closed-loop operation that suppresses contrast loss under high-dynamic conditions.

Finally, we discuss the noise performance achievable by a closed-loop quantum inertial sensor.  
In the simulation, shot noise was applied to the final atomic flux, and the result was multiplied by the scale factor to evaluate the velocity random walk (VRW) for acceleration and the angular random walk (ARW) for angular velocity.
In this simulation, the alternating phase shift of Raman beam~B was set to 
$\Delta \Phi = \pi/2$.
For the experimental parameters assumed in this simulation, a VRW of \SI{3}{\micro m / s^2 / \sqrt{Hz}} and an ARW of \SI{15}{\micro deg / \sqrt{h}} were obtained.  

The VRW and ARW obtained in the simulation are also analytically given by the following expressions:
\begin{align}
\mathrm{VRW} &= \frac{1}{C \, \frac{\partial \phi_a}{\partial a} \, \sqrt{\dot{N}}}, \\
\mathrm{ARW} &= \frac{1}{C \, \frac{\partial \phi_\Omega}{\partial \Omega} \, \sqrt{\dot{N}}},
\end{align}
where $C$ is the interference contrast and $\dot{N}$ is the atomic flux.  
In Eqs.~(3) and (4), when the atomic velocity $v$ is represented by the most probable velocity $v_{\mathrm{mp}}$, the values are 
\SI{3}{\micro m / s^2 / \sqrt{Hz}} and an ARW of \SI{15}{\micro deg / \sqrt{h}}, respectively, which are in perfect agreement with the simulation results.  
In the analytical calculation, it is also taken into account that the use of counter-propagating atomic beams effectively doubles the atomic flux.

Note that in the sensitivity estimation, we use the total atomic flux emitted from the oven for both counter-propagating atomic beams, $1.6 \times 10^{11}~{\rm atoms/s}$.
This corresponds to an idealized case in which all atoms contributing to the interferometer participate in the measurement.
In a practical implementation, the effective atomic flux may be reduced by several factors, such as the finite velocity acceptance determined by the Raman transition linewidth, spatial inhomogeneity of the laser intensity leading to imperfect pulse areas, and other technical issues.
The sensitivity values presented here should therefore be regarded as best achievable limits under the assumed experimental conditions.

Among the highest-performance mechanical accelerometers widely used in inertial navigation systems is the quartz pendulum accelerometer, in which the apparent force applied to a quartz pendulum is balanced by an electromagnetic force in a closed-loop configuration; a representative VRW for this device is \SI{2}{\micro m / s^2 / \sqrt{Hz}}~\cite{beitia_clifford_fell_loisel_2015}.
Furthermore, in the field of gyroscopes for inertial navigation, fiber-optic gyroscopes have recently surpassed the performance of ring laser gyroscopes, with ARW below \SI{100}{\micro deg / \sqrt{h}}~\cite{paturel_one_2014}.
These results indicate that the digital closed-loop quantum inertial sensor has the potential to surpass state-of-the-art performance in both acceleration and angular velocity measurements.

\section{Summary}
In this paper, we have demonstrated an innovative approach to a thermal atomic beam interferometer, a system expected to offer both fast response and high sensitivity.
Our method applies the ``digital closed-loop'' concept—originally developed in the FOGs—to thermal atomic beam interferometers.
This approach not only enables the simultaneous measurement of acceleration and angular velocity, which has conventionally been challenging, but also significantly expands the dynamic range for both measurements.
Under compact and realistic experimental parameters, the anticipated VRW and ARW values surpass those of sensors for the modern state-of-the-art inertial navigation system, while the dynamic range covers gravitational acceleration as well as high dynamics on the order of several tens of degrees per second.
This demonstrates that the proposed method has the potential to be applied not only to low-dynamic vehicles, such as autonomous underwater vehicles, but also to a wide range of platforms for inertial navigation, including automobiles, aircraft, and spacecraft.
Moreover, owing to the inherent characteristics of the digital closed loop, synchronizing multiple thermal atomic beam interferometers is straightforward—an extremely important advantage for constructing inertial navigation systems, which must handle motion not limited to one-dimensional translation or single-axis rotation, but occurring in the full three-dimensional space.
Therefore, the present proposal is regarded as providing a key protocol for the realization of a full quantum-based inertial navigation system. We are presently engaged in developing such a system based on this approach.

\section*{Acknowledgments}
We are grateful to Yuichiro Kamino for valuable discussions. This work was supported by JST, Japan, under Grant Numbers JPMJMI17A3 and JPMJPF2015.

\section*{Data availability}
The data that support the findings of this article are openly available~\cite{dataset}.

\appendix
\section{}
Here, we present the equations from the analytical expressions for the two-photon Raman process derived by B. Young \textit{et al.}~\cite{youngPrecisionAtomInterferometry1997} that were used in carrying out the simulations in this paper, and describe the equation expansions employed in the interference calculations.
The atomic state is represented as a superposition of two states coupled via the Raman transition:
\begin{equation}
\begin{aligned}
|\Psi(t)\rangle &= C_{e,\,p+\hbar k_{\mathrm{eff}}}(t) 
\, e^{-i\left( \omega_{e} + \frac{|p+\hbar k_{\mathrm{eff}}|^{2}}{2m\hbar} \right)t} 
\, |e,\,p+\hbar k_{\mathrm{eff}}\rangle \\
&\quad + C_{g,\,p}(t) 
\, e^{-i\left( \omega_{g} + \frac{|p|^{2}}{2m\hbar} \right)t} 
\, |g,\,p\rangle
\end{aligned}
\end{equation}
where $C_{e,\,p+\hbar k_{\mathrm{eff}}}(t)$ and $C_{g,\,p}(t)$ are the amplitude coefficients of the excited state with a momentum kick and the ground state without a momentum kick, respectively.
When Raman beams with equal intensity are applied over the interval $t_0 \le t \le t_0 + \tau$, the amplitude coefficients take the form
\begin{equation}
\begin{aligned}
&C_{e,\,p+\hbar k_{\mathrm{eff}}}(t_0 + \tau) \\
&= e^{-i\left(\Omega_e^{\mathrm{AC}} + \Omega_g^{\mathrm{AC}}\right)\tau/2} \,
e^{-i \delta_{12} \tau / 2} \\
&\quad \times \Bigg\{
C_{e,\,p+\hbar k_{\mathrm{eff}}}(t_0)
\left[ \cos\left( \frac{\Omega'_r \tau}{2} \right)
- i \cos\Theta \, \sin\left( \frac{\Omega'_r \tau}{2} \right) \right] \\
&\qquad
+ C_{g,\,p}(t_0) \,
e^{-i(\delta_{12} t_0 + \phi_{\mathrm{eff}})}
\left[ -i \sin\Theta \, \sin\left( \frac{\Omega'_r \tau}{2} \right) \right]
\Bigg\}
\end{aligned}
\label{C_e}
\end{equation}
\begin{equation}
\begin{aligned}
&C_{g,\,p}(t_0 + \tau) \\
&= e^{-i\left(\Omega_e^{\mathrm{AC}} + \Omega_g^{\mathrm{AC}}\right)\tau/2} \,
e^{i \delta_{12} \tau / 2} \\
&\quad \times \Bigg\{
C_{e,\,p+\hbar k_{\mathrm{eff}}}(t_0) \,
e^{i(\delta_{12} t_0 + \phi_{\mathrm{eff}})}
\left[ -i \sin\Theta \, \sin\left( \frac{\Omega'_r \tau}{2} \right) \right] \\
&\qquad
+ C_{g,\,p}(t_0)
\left[ \cos\left( \frac{\Omega'_r \tau}{2} \right)
+ i \cos\Theta \, \sin\left( \frac{\Omega'_r \tau}{2} \right) \right]
\Bigg\}.
\end{aligned}
\label{C_g}
\end{equation}
Here, $\phi_{\mathrm{eff}}$ represents the Raman optical phase and corresponds to $\phi_{\mathrm{A}}$, $\phi_{\mathrm{B}}$, and $\phi_{\mathrm{C}}$ in the main text.
When $\Omega_g$ and $\Omega_e$ represents the single-photon Rabi frequencies for the 
$\lvert g \rangle \leftrightarrow \lvert i \rangle$ and 
$\lvert e \rangle \leftrightarrow \lvert i \rangle$ transitions, respectively,
and the two-photon Rabi frequency is denoted as $\Omega_{\mathrm{eff}}$,
we define
\begin{equation}
\Omega_e^{\rm{AC}} \equiv \frac{|\Omega_e|^2}{4\Delta}, \quad
\Omega_g^{\rm{AC}} \equiv \frac{|\Omega_g|^2}{4\Delta}
\end{equation}
\begin{equation}
\delta^{\rm{AC}} \equiv \Omega_e^{\rm{AC}} - \Omega_g^{\rm{AC}}
\end{equation}
\begin{equation}
\Omega_r' \equiv \sqrt{\Omega_{\mathrm{eff}}^2 + \left( \delta_{12} - \delta^{\rm{AC}} \right)^2}
\end{equation}
\begin{equation}
\sin\Theta = \frac{\Omega_{\mathrm{eff}}}{\Omega_r'}, \quad
\cos\Theta = -\frac{\delta_{12} - \delta^{\rm{AC}}}{\Omega_r'}.
\end{equation}
\begin{figure}[htb]
\vspace{-0.0cm}
    \centering
    \includegraphics[scale=0.33]{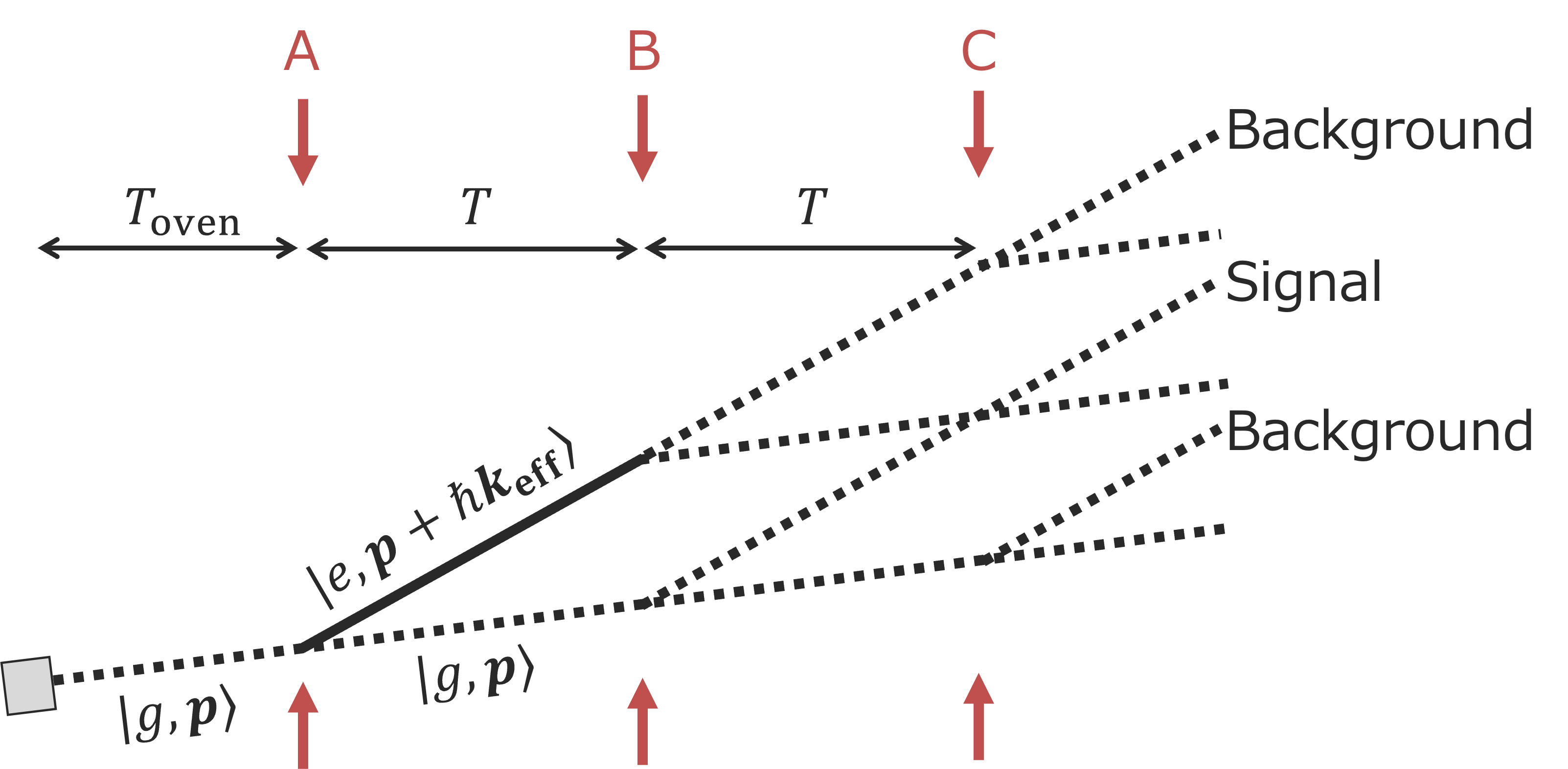}
    \caption{Atom interferometer using a right-propagating atomic beam. 
$T_{\mathrm{oven}}$: time from the oven to Raman beam~A. 
$T$: time between Raman beams. 
Population in the $\lvert e \rangle$ state is measured via atomic fluorescence, 
with the interferometric contribution (\textit{Signal}) and non-interferometric component (\textit{Background}).
}
    \label{fig_interference_calc}
\vspace{-0.0cm}
\end{figure}

Here, $\Theta$ satisfies $0 \leq \Theta \leq \pi$. In the digital closed-loop scheme, four types of interferometers are constructed.  
In all cases, a deviation of the atomic velocity from $v_{\mathrm{mp}}$ causes the pulse area to deviate from the $\pi/2 - \pi - \pi/2$ condition, leading to complex trajectories, as shown in Fig.~\ref{fig_interference_calc}.
Since the atomic population in the $\lvert e\rangle$ state is monitored via fluorescence detection, the measured signal corresponds to the sum of the closed-interferometer contribution (\textit{Signal}) and the non-interferometric, open-path component (\textit{Background}).  
By appropriately accounting for these components, it becomes possible to simulate both the interferometric contrast and the interference signal under applied acceleration and rotation.  

As an example, we calculate the probability amplitude for the portion indicated by the solid line in the figure.  
With an appropriate laser intensity, the AC Stark shift can be cancelled, and thus it is neglected here.  
From Eqs.~(\ref{C_e}) and (\ref{delta12A}), the probability amplitude $\eta$ is given by  
\begin{align}
\eta = & e^{-i\alpha\tau/2} \, e^{-i(\beta+\phi_{\rm A})} 
   \left[ -i \sin\Theta^{(\rm{A})} \, \sin\!\left( \frac{\Omega_{r}^{\prime (\rm{A})}\tau}{2} \right) \right], \label{eq:ampA}
\end{align}
where
\begin{align}
\alpha &= \delta + \gamma T_{\mathrm{oven}} 
- k_{\mathrm{eff}}\!\left( \Omega L + a T_{\mathrm{oven}} + \Delta v \sin\theta \right), \\
\beta &= \delta T_{\mathrm{oven}} + \gamma T_{\mathrm{oven}}^2 \notag\\
&\quad - k_{\mathrm{eff}}\!\left( \Omega L T_{\mathrm{oven}} 
+ \frac{1}{2} a T_{\mathrm{oven}}^2 
+ \Delta v \sin\theta \, T_{\mathrm{oven}} \right).
\end{align}

\par
\phantom
\newline
\phantom
\newline
\phantom
\newline
\phantom
\newline
\phantom
\newline

\FloatBarrier

\bibliographystyle{apsrev4-2}
\bibliography{digital_closed-loop_marged}

\end{document}